\documentclass{aastex62}

  \newcommand{\dprime}{{\prime\prime}}
  
  \newcommand{\noise}{{\mathcal N}}

  \usepackage{amsmath}
  \usepackage{graphicx}
  \usepackage[utf8]{inputenc}

\renewcommand{\arraystretch}{1.5}
  \graphicspath{{./}{figures/}}

  \submitjournal{ApJ}

  \shorttitle{Rossby waves in the Sun}

\begin{document}


\title{Properties of solar Rossby waves from normal mode coupling and characterizing its systematics}

  \correspondingauthor{Krishnendu Mandal}
  \email{krishnendu.mandal@tifr.res.in}
  
   \author[0000-0003-3067-288X]{Krishnendu Mandal}
  
  \author[0000-0003-2896-1471]{Shravan Hanasoge}
 

\affiliation{Department of Astronomy \& Astrophysics,
  Tata Institute of Fundamental Research, Mumbai 400005}

  \begin{abstract}
    Rossby waves play an important role in mediating the angular momentum of rotating spherical fluids, creating weather on Earth and tuning exoplanet orbits in distant stellar systems \citep{ogilvie14}. Their recent discovery in the solar convection zone provides an exciting opportunity to appreciate the detailed astrophysics of Rossby waves. Large-scale Rossby waves create subtle drifts in acoustic oscillations in the convection zone, which we measure using helioseismology to image properties of Rossby waves in the interior. We analyze $20$ years of space-based observations, from $1999$ to $2018$, to measure Rossby-mode frequencies, line-widths and amplitudes. Spatial leakage affects the measurements of normal-mode eigenfunction coupling (which we refer to as ``normal-mode coupling" in this paper) and complicates the analysis of separating out specific harmonic degree and azimuthal number of features on the Sun. Here we demonstrate a novel approach to overcome this difficulty and test it by performing synthetic tests. We find that the root-mean-square velocity of the modes is of the order of $0.5$ m/s at the surface.   
   
  \end{abstract}
  \keywords{helioseismology, waves --- 
  miscellaneous --- catalogs --- surveys}

 \section{Introduction} \label{sec:intro}
Rossby waves were first detected as large-scale weather patterns in Earth's atmosphere \citep{rossby_1939} and subsequently in the ocean \citep{chelton96}. Theoretical analysis \citep{Papa1978, unno_1989,provost_81,smeyers_81,saio_82} suggested that rotating spherical fluids should show the presence of Rossby waves (commonly known as r-modes oscillations in the astrophysical context), in which the Coriolis force acts as a restoring force. The Sun, a differentially rotating spherical fluid, satisfies this condition. Long-term high-quality observations have encouraged several attempts in the past to detect Rossby waves \citep{Kuhn2000,ulrich2001,sturrock2015}. In all these earlier studies, the dispersion relation was not measured, a critical quantity in the identification of the governing physics. Recently, \citet{gizon18}, using granulation tracking and the helioseismic technique of ring diagrams \citep{hill88} measured a Rossby-dispersion relation, thereby unambiguously detecting sectoral modes (where the azimuthal order of the mode is the same as its harmonic degree) in the subsurface through analyses of six years of data from the Helioseismic and Magnetic Imager \citep[HMI:][]{hmi} onboard the Solar Dynamics Observatory. Subsequently, \citet{liang_2018} and \citet{hanasoge19}, using two different methods, time-distance helioseismology \citep{duvall} and normal mode coupling respectively, further validated the original detection.  All the studies mentioned above have focused on Rossby waves in the solar surface and near-surface regions. Since Rossby waves exist in spherical rotating fluids, it can also in principle exist in solar atmosphere. \cite{mcintosh2017} using the tracking of coronal bright points has found evidence of Westward propagating large-scale Rossby-like waves in the solar corona. It may be useful to characterize the connection between solar internal Rossby waves and the measurements of \citet{mcintosh2017}. One important difference between them is that solar internal Rossby waves are retrograde waves whereas Rossby like waves detected in solar corona by \citet{mcintosh2017} are prograde propagating. \citet{zaqa_mag_rossby2010,zaqa_QBO_2010,zaqa_2015,dikpati2018,dikpati2018a} studied magnetized Rossby waves in a magneto-hydrodynamic shallow water model of the solar tachocline in order to understand quasi-biennial oscillations and Rieger-type periodicities \footnote{Solar activity follows a dominant $11$-year cycle. Two other timescales have also been observed in solar activity indicators, the first presenting timescales of around $150$-$160$ days - commonly known as Rieger-type periodicity and the second following an approximately-$2$ year periodicity, known as quasi-biennial oscillations.} in solar activity. It is therefore important to characterize the properties of internal solar Rossby waves from observations in order to determine how it affects solar internal dynamics.\par

Normal-mode coupling, a seismic technique with an illustrious history in geophysics \citep[see, e.g.][]{DT98}, has seen limited adoption in helioseismology. \citet{woodard_89} first described distortion of eigenfunctions of the Sun due to latitudinal differential rotation. This method has subsequently been used by several authors e.g. \citet{lavely92,woodard06,roth_stix_2008,schad,schad_2013,woodard13,woodard14,woodard16, hanasoge17, hanasoge18} in various studies ranging from meridional circulation to convection. \citet{roth2003,woodard16, hanasoge18,hanasoge17_etal} have showed how this method may be used to glean information about time-varying, non-axisymmetric features in the Sun. The fundamental measurement comprises  cross-correlated Fourier coefficients of the wavefield, $\langle\phi^*_{\ell^{\prime} m^\prime}(\omega^\prime)\phi_{\ell m}(\omega)\rangle$, where angular brackets denote ensemble averaging, $\phi$ the line-of-sight Doppler velocity, $\ell$ and $m$ harmonic degree and azimuthal order respectively, and $\omega$ the temporal frequency. In the absence of perturbations, the expected value of this correlation is precisely obtained by considering leakage and by modelling modes as independently and identically distributed \footnote{If an outcome of a random sampling is independent of the random variables that came before it, then we call the process independent and identically distributed in short i.i.d} random processes. Depending on the structure of the perturbations, specific spatio-temporal wavenumbers show enhanced power above the background systematic floor. This constitutes a ``detection" of a perturbation at those spatial and temporal scales. The difference between temporal frequencies $\sigma = \omega^{\prime}-\omega$ captures information about the perturbation time scale, whereas the difference between harmonic degrees, $\ell^{\prime}-\ell$, and azimuthal number, $m^\prime-m$  carry information about the length scale, and toroidal and poloidal nature of flows. \citet{hanasoge19} used this formalism to detect Rossby waves in the Sun with the first $2$ years of HMI data, thereby validating the methodology.  In this work, we have extended the analysis of \citet{hanasoge19} with $12$ and $8$ years of solar oscillation data from Michelson Doppler Imager (MDI) onboard the Solar and Heliospheric Observatory and HMI respectively to estimate frequencies, line-widths and amplitudes of Rossby modes and compare them with parameters obtained from earlier studies by \citet{gizon18, liang_2018}. Because we do not observe the entire Sun, spatial and temporal leakage affects our measurements. We perform several synthetic tests in order to characterize the effect of leakage, in order to be able to place faith in our inferences, e.g. to recover the depth profile of Rossby modes accurately from a synthetic measurement with added noise. This will help us further determine depth profiles of Rossby waves in the convection zone, which are still unknown and are active areas of research. 

 \section{\label{sec:data}Data Analysis}
 The mode-coupling measurement procedure for Rossby waves is described in detail in \citet{hanasoge19} and \citet{hanasoge18}. We follow same notations as \citet{hanasoge18} for convenience. The raw data are global time series of line-of-sight Doppler velocity projected on to spherical harmonics, i.e. $\phi_{\ell m}$ from MDI and HMI. We calculate the cross-correlation function $\langle\phi^*_{\ell m}(\omega)\phi_{\ell m+t}(\omega+\sigma)\rangle$. Analyzing this quantity at each frequency, $\omega$, is less tractable, and we therefore define B-coefficients as a linear least-square fit to the raw wavefield correlations,
 
 \begin{equation}
     B^{\sigma}_{st}(n,\ell)=\frac{\sum_{m,\omega}H^\sigma_{\ell mst}(\omega)\phi^{*}_{\ell m}(\omega)
     \phi_{\ell m+t}(\omega+\sigma+t\Omega)}{\sum_{m,\omega}\vert H^\sigma_{\ell mst} \vert^{2}},\label{eq:B_coeff}
 \end{equation}
 where $H$ is a weight function defined in \citet{hanasoge19} and \citet{hanasoge18} 
 \begin{equation}
     H^\sigma_{\ell m s t}(\omega) = -2\omega(-1)^{m+t}\sqrt{2s+1}\begin{pmatrix} \ell & s & \ell \\ -(m+t) & t & m\end{pmatrix}
L_{\ell m}^{\ell m}L_{\ell	m+t}^{\ell	m+t}N_\ell(R^{\omega*}_{\ell m} | R^{\omega+\sigma+t\Omega}_{\ell m+t}|^2 +
|R^{\omega}_{\ell m}|^2  R^{\omega+\sigma+t\Omega}_{\ell m+t} ),\label{eqH}
 \end{equation}
 where the first term on the right hand side of Equation~(\ref{eqH}) is a Wigner-$3j$ symbol and $L^{\ell^\prime m^\prime}_{\ell m}$ is the leakage matrix, which describes how spatial windowing in the data, i.e. arising from our limited vantage of the Sun, causes ``leakage" from mode $(\ell,m)$ to another mode $(\ell^\prime,m^\prime)$. $B$-coefficients are calculated for all identified radial orders, $(n)$ and harmonic degrees in the range $\ell\in[10,180]$. $R^{\omega}_{\ell m}$ describes the power spectrum of a mode (labelled using three quantum numbers $n,\ell,m$) \citep{anderson_1990,Duvall_harvey_93},
 \begin{equation}
     R^{\omega}_{\ell m}=\frac{1}{(\omega_{n \ell m}-i\Gamma_{n\ell}/2)^2-\omega^{2}},
 \end{equation}
 where $\omega_{n\ell m}$ is an eigenfrequency and $\Gamma_{n\ell}$ is the full width at half maximum. We use observed values, i.e. obtained through fits of spectra by the global-mode pipeline, for these parameters.  We assume that errors in measurements of $\langle\phi^*_{\ell m}(\omega)\phi_{\ell m+t}(\omega+\sigma)\rangle$ are uncorrelated, have equal variances and zero expectation value. This allows us to invoke Gauss-Markov theorem which ensures that B-coefficients estimated from Equation \ref{eq:B_coeff} as an unregularized least square estimate of the cross correlation measurement are unbiased. Differential rotation advects features on the Sun at a variety of different speeds. In order to appropriately follow these perturbations, we adopt a co-rotating frame, i.e. apply a tracking rate. The temporal frequencies of the perturbation in the co-rotating frame, $\sigma$, will be transformed to $\sigma+t\Omega$ in the inertial frame, where $\Omega$ is rotation frequency of the Sun. A variety of tracking rates may be chosen since the Sun is differentially rotating: here, we choose $453$ nHz, which is the value at the equator. 
 We extend the frequency range from our earlier work \citep{hanasoge19} to $\sigma\in[0,200]$ nHz in order to appreciate the spectra of the perturbations better.
 
 \section{Inversion}
  We assume Rossby waves are sufficiently well described by a toroidal flow, allowing it to be expressed as following 
  \begin{equation}
     u^\sigma(r,\theta,\phi)=\sum_{st}w^\sigma_{st}(r)\,\hat{\mathbf{r}}\times{\boldsymbol{\nabla}}_h Y_{st}(\theta,\phi),\label{eq:flow}
  \end{equation}
  where $(r,\theta,\phi)$ and $(\hat{\mathbf{r}},\hat{\bm{\theta}},\hat{\bm{\phi}})$ are radius, co-latitude, longitude and corresponding unit vectors respectively, $\bm{\nabla}_h$ is the horizontal covariant derivative and $Y_{st}$ is a spherical harmonic of degree $s$ and azimuthal-order $t$. $w^\sigma_{st}(r)$ determines the depth variation of Rossby waves. 
  
  We observe only the near side of the Sun, the angular extent that appears in the field of view of the telescope. This results in spatial and temporal leakage in the measurements (see section~\ref{sec:leakage}). B-coefficients as estimated in section~(\ref{sec:data}) will encounter leakage from neighbouring harmonic degrees and azimuthal number \citep[see Equation 26 of][]{hanasoge18}
  \begin{equation}
      B^\sigma_{st}(n,\ell)=N^\sigma_{\ell st}\sum_{\ell^\prime,\ell^{\prime\prime},m,m^\prime,s^\prime
      ,t^\prime,\omega}L_{\ell m}^{\ell^\prime,m^{\prime}}L_{\ell m+t}^{\ell^{\prime\prime}m^\prime+t^\prime}\gamma_{tm}^{\ell s\ell}H^{\sigma *}_{\ell\ell mt}
      \gamma^{\ell^{\prime\prime}s^\prime\ell^\prime}_{t^\prime m^\prime}H^{\sigma}_{\ell^{\prime}\ell^{\prime\prime}m^{\prime}t^{\prime}}b^\sigma_{s^\prime t^\prime}(\ell^\prime,\ell^{\prime\prime}),\label{eq:B_leak}
  \end{equation}
  
  where  
  \begin{equation}
  N^\sigma_{\ell st}=\frac{1}{\sum_{m,\omega}\vert\gamma_{tm}^{\ell s\ell}H^{\sigma}_{\ell\ell mt}\vert^2},
  \label{eq:N_sigma}
  \end{equation}
  and
  \begin{equation}
      b^{\sigma}_{st}(\ell,\ell^{\prime})=f_{\ell^{\prime}-\ell,s}\int_{\odot}dr w_{st}^{\sigma}(r)\mathcal{K}_{n \ell}(r).\label{B_no_leak}
  \end{equation}
  $B$ and $b$-coefficients refer to different quantities and are only in special cases (as discussed below) identical.  $B^\sigma_{st}$ and $ b^{\sigma}_{st}$ are also functions of radial order, $n$. In this work we consider the coupling between modes with same radial order and therefore omit it in the expression for notational brevity. $\mathcal{K}_{n\ell}$ is the sensitivity kernel for the mode $(n,\ell)$ and $f_{\ell^{\prime}-\ell,s}$ is obtained from an asymptotic analysis of the kernels \citep{vorontsov11, hanasoge18, woodard14}. This asymptotic kernel is only valid when $s\ll \ell$ or $s\ll \ell^\prime$ \citep[see for example][]{brussaard_1957,vorontsov11}. \citet{hanasoge18} have shown a comparison between kernels obtained from asymptotic analysis and the exact expression (Figure 10 of that paper) and found that the comparison gets slightly worse when $s\approx \ell$ but is otherwise very accurate when $s<\ell$. In this work, we search for coupling between p-modes with harmonic degree in the range $\ell \in [10,180]$, and are interested in Rossby modes with harmonic degree $s\leq 20$, thereby justifying the use of asymptotic kernels. In case of full-sphere observations, we show in appendix \ref{sec:appendix2} that the measured B-coefficients from Equation~(\ref{eq:B_coeff}) will be reduced to Equation~(\ref{B_no_leak}), a simple relation with which to invert for the velocity profile $w^{\sigma}_{st}$.

  Inverting Equation~(\ref{eq:B_leak}) is complicated in the general case, as explained in \citet{hanasoge18}  due to leakage contributions from neighbouring modes into the target measurement channel. 
  For solar Rossby waves, we know from earlier studies by \citet{gizon18,liang_2018,hanasoge19} that non-sectoral modes $(s\neq\vert t \vert)$ are either absent or appear at amplitude levels well below the detection threshold. Due to this condition, Equation \ref{eq:B_leak} would be simplified after substituting $b^\sigma_{st}\approx\delta_{s,-t}b^\sigma_{s -s}$
  \begin{equation}
      B^\sigma_{s, -s}(n,\ell)=N^\sigma_{\ell st}\sum_{\ell^\prime,\ell^{\prime\prime},m,m^\prime,s^\prime
      }L_{\ell m}^{\ell^\prime,m^{\prime}}L_{\ell m-s}^{\ell^{\prime\prime}m^\prime-s^\prime}\gamma_{-s \,m}^{\ell s\ell}H^{\sigma *}_{\ell\ell m\,-s}
      \gamma^{\ell^{\prime\prime}s^\prime\ell^\prime}_{-s m^\prime}H^{\sigma}_{\ell^{\prime}\ell^{\prime\prime}m^{\prime}-s^\prime}b^\sigma_{s^\prime ,-s^\prime}(\ell^\prime,\ell^{\prime\prime}).\label{eq:B_leak_ns}
  \end{equation}
  Although Equation \ref{eq:B_leak_ns} is much simpler than Equation \ref{eq:B_leak}, we still need to show that we can separate out desired specific spatial scales using our measurement. We follow two approaches as discussed below for that. In our first approach, we do not consider leakage and invert Equation~(\ref{B_no_leak}), assuming $b^\sigma_{s,-s}$ is the same as $B^\sigma_{s,-s}$, as considered in our earlier work \citep{hanasoge19}. This approximation will only hold if there is no leakage, i.e., $L^{\ell^\prime m^\prime}_{\ell m}=\delta_{\ell \ell^\prime}\delta_{m m^\prime}$. The approximation affects our inferences and to quantify this, we perform synthetic tests. In our second approach we consider leakage and a discussion of which is presented in section ~(\ref{sec:syn_inv}). We choose two inversion techniques, Optimally Localized Averaging, \citep[OLA;][]{backus1967} and Regularized Least Square, (RLS) for this work.   
  

 \subsection{\label{OLA}OLA}
In OLA, the inverted flow profile at depth $r_0$ is written as a linear combination of all the measurements as following,
   \begin{equation}
      w^{\sigma}_{st}(r_0)=\sum_{n\ell} \alpha_{n\ell;r_0}B^{\sigma}_{st}(n,\ell),\label{w_alpha}
  \end{equation}
  where $\alpha$ needs to be determined by minimizing the following misfit function
  \begin{equation}
      \chi=\int_{\odot}dr\,12(r-r_0)^2\left(\sum_{n,\ell}\alpha_{n\ell;r_0}f_{0,s}K_{n\ell}(r)
      \right)^2+\lambda\sum_{n\ell}\mathcal{N}_{n\ell}\alpha_{n\ell;r_0}^2.\label{eq:OLA_chi}
  \end{equation}
  $\lambda$ is the regularization parameter. $\sum_{n,\ell}\alpha_{n\ell;r_0}f_{0,s}K_{n\ell}$ determines the averaging kernel at depth $r_0$. Weight function $(r-r_0)^2$ in the first term of Equation~(\ref{eq:OLA_chi}) ensures that the averaging kernel obtained after minimizing Equation~(\ref{eq:OLA_chi}) is large valued at depth $r_0$ and small elsewhere. We can choose to perform the inversion separately for each frequency bin or once for all frequency bins. In this technique, we choose the latter approach: $\noise_{n\ell}$ is obtained from $\mathcal{N}^{\sigma}_{n\ell}$ after summing over all the frequency bins, $\sigma$, where $\mathcal{N}^{\sigma}_{n\ell}$ is the diagonal component of the noise covariance matrix. In the next section we perform an inversion for each frequency bin separately using the RLS method. 
   
 \subsection{\label{RLS} RLS}
 In this method, we expand the flow profile in the B-spline basis 
 \begin{equation}
     w^{\sigma}_{st}(r)=\sum_{k}\beta^{\sigma}_{st}B_k(r),\label{RLS_exp}
 \end{equation}
 where $B_k$ is the B-spline basis function of order three. We choose total $50$ knots for the inversion up to depth $0.1 R_\odot$. We determine $\beta$ by minimizing the following misfit function
 \begin{equation}
     \chi=\sum_{n,\ell}\frac{\left( B^\sigma_{st}(n,\ell)-f_{0,s}\int_{\odot}dr w^{\sigma}_{st}(r)\mathcal{K}_{n \ell}(r)\right)}{\mathcal{N}^{\sigma}_{n\ell}}^2+
     \lambda\sum \left(\frac{d^2 w^{\sigma}_{st}}{d r^2}\right)^2,\label{RLS_chi}
 \end{equation}
 where we consider second-derivative smoothing where $\lambda$ is the regularization parameter. The Equation~(\ref{RLS_chi}) may be minimized if we solve the following system of equations
 \begin{equation}
     \frac{1}{\mathcal{N}^{\sigma}_{n\ell}}f_{0,s}\int_{\odot}dr w^\sigma_{st}(r)\mathcal{K}_{n \ell}(r)=\frac{1}{\mathcal{N}^{\sigma}_{n\ell}}B^\sigma_{st}(n,\ell),
 \end{equation}
 \begin{equation}
     \lambda \frac{d^2 w^{\sigma}_{st}(r_k)}{d r^2}=0,
 \end{equation}
where $r_k$ denotes points on the radial grid. As opposed to earlier methods, we must perform inversions for separate frequency bins in RLS. 
 
 \section{Results}
 
\subsection{\label{sec:leakage}Modelling leakage}
\begin{figure}
\begin{centering}
\includegraphics[scale=0.5]{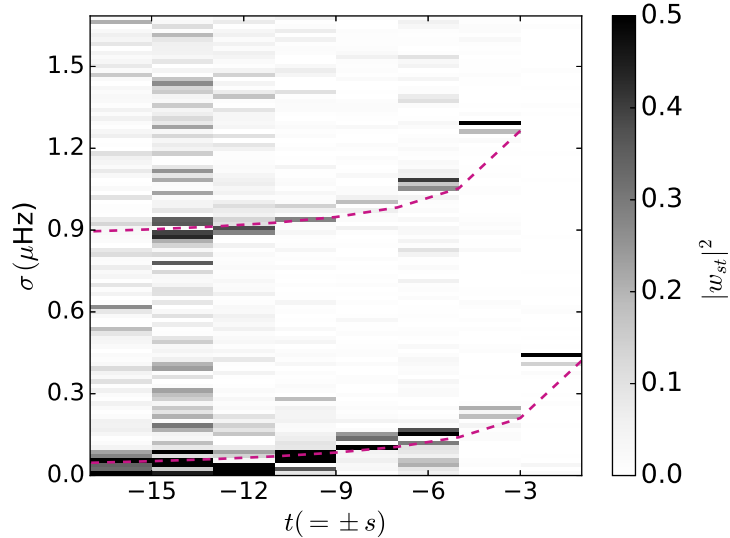}
\caption{\label{fig:leakage_data}{Rossby modes of harmonic degree $s$ and frequency $\sigma_s$ leak into degree $s+2$ with frequency $\sigma_{s}+2\Omega$. Red dashed lines in the lower and upper parts of the figure show the classical Rossby-wave dispersion relation (Equation \ref{dispersion}) and leakage of those modes into higher frequencies.}}
\par\end{centering}
\end{figure}

\citet{hanasoge19} considered a frequency analysis $\sigma$ in the range $(0.0,0.5)\,\mu$Hz. In this work, we expand the range of $\sigma$ up to $2\,\mu$Hz and show in Figure~(\ref{fig:leakage_data}) that Rossby modes leak into higher frequencies, much as observed by \citet{liang_2018}. The sectoral mode of harmonic degree $s$ and corresponding frequency $\sigma_s = 2\Omega/(s+1)$ leaks into a harmonic degree $s+2$ at frequency $\sigma_s+2\Omega$. Spatial windowing of the rotating Sun results in simultaneous spatial and temporal leakage, which we verify through a synthetic test. \par
We consider sectoral modes of Rossby waves for odd harmonic degrees. Rossby waves in the rotating Sun are multiplied by a spatial window function - which is unity for the visible portion of the disk, and otherwise zero. We perform spherical harmonic and temporal Fourier transforms in order to estimate $v_{s}(\omega)$. Since we only detect sectoral retrograde modes, i.e. $t = -s$, we use both $t$ and $s$ equivalently to denote Rossby waves. Analytical calculations suggest (see Equation \ref{vs_delta} in Appendix~\ref{sec:appendix}) that the observed $v_{s}(\omega)$ contains contributions from neighbouring modes $s^\prime$ due to leakage $L_{s}^{s^\prime}$,
    
\begin{equation}
    v_{s}(\omega)=\sum_{s
    ^\prime}L_{s}^{s^\prime}w_{s^\prime }\delta(\omega-(\sigma_{t^\prime}+t^\prime\Omega)).\label{temporal_leakage}
\end{equation}
The Delta function is invoked assuming that Rossby modes have power close to the classical dispersion frequency
\begin{equation}
    \sigma_t=\frac{2\Omega}{2\vert t\vert+1},\label{dispersion}  
\end{equation}
where $\sigma_t$ is the frequency of the Rossby mode with azimuthal number $t$ in a co-rotating frame with rotation frequency $\Omega$. From Equation~(\ref{temporal_leakage}) we see that neighbouring modes $s^\prime$ will corrupt the measurements of our desired mode $s$ by an amount that depends on the value of leakage $L_{s}^{s^\prime}$.   \par

\begin{figure}
\begin{centering}
\includegraphics[scale=0.45]{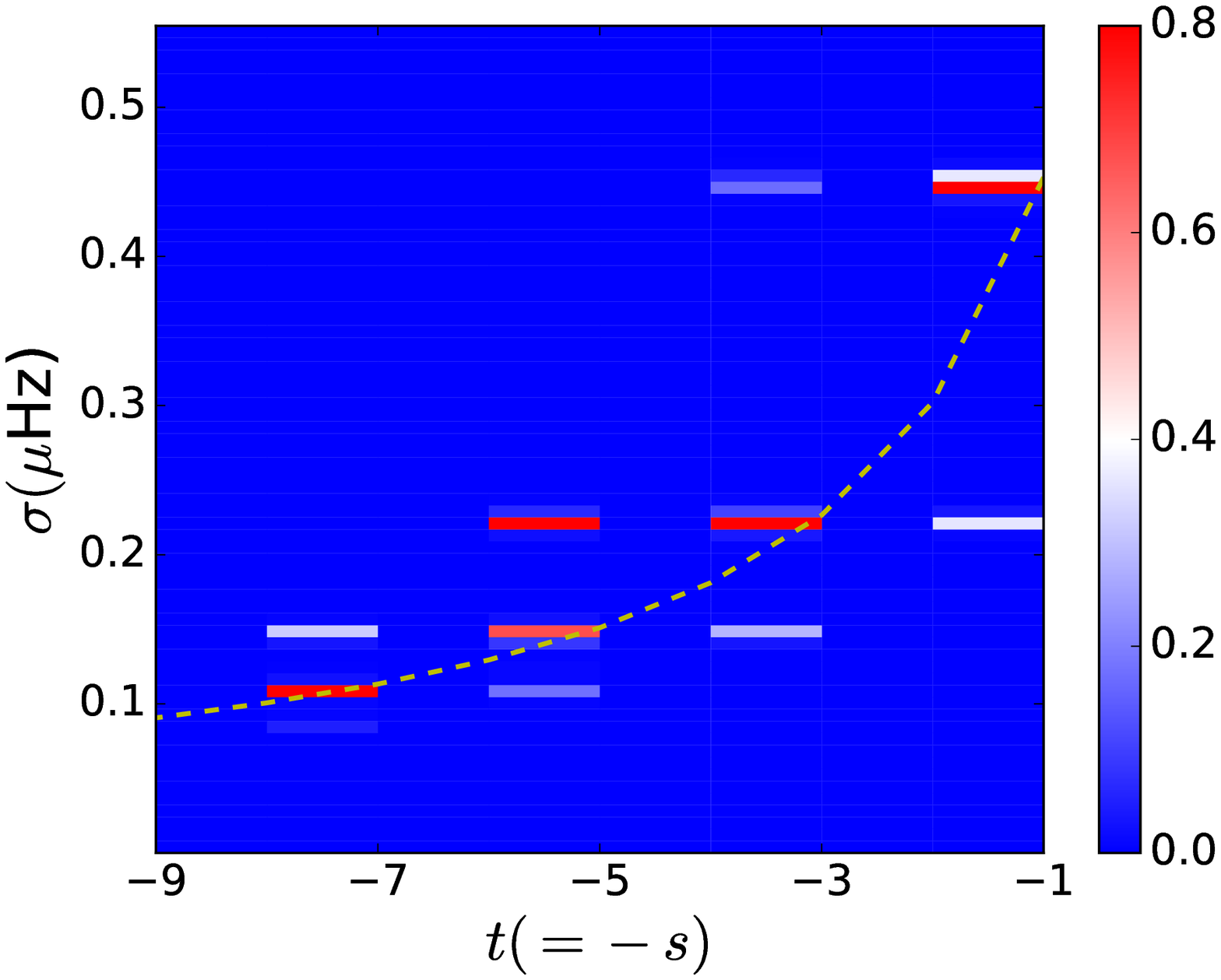}\includegraphics[scale=0.45]{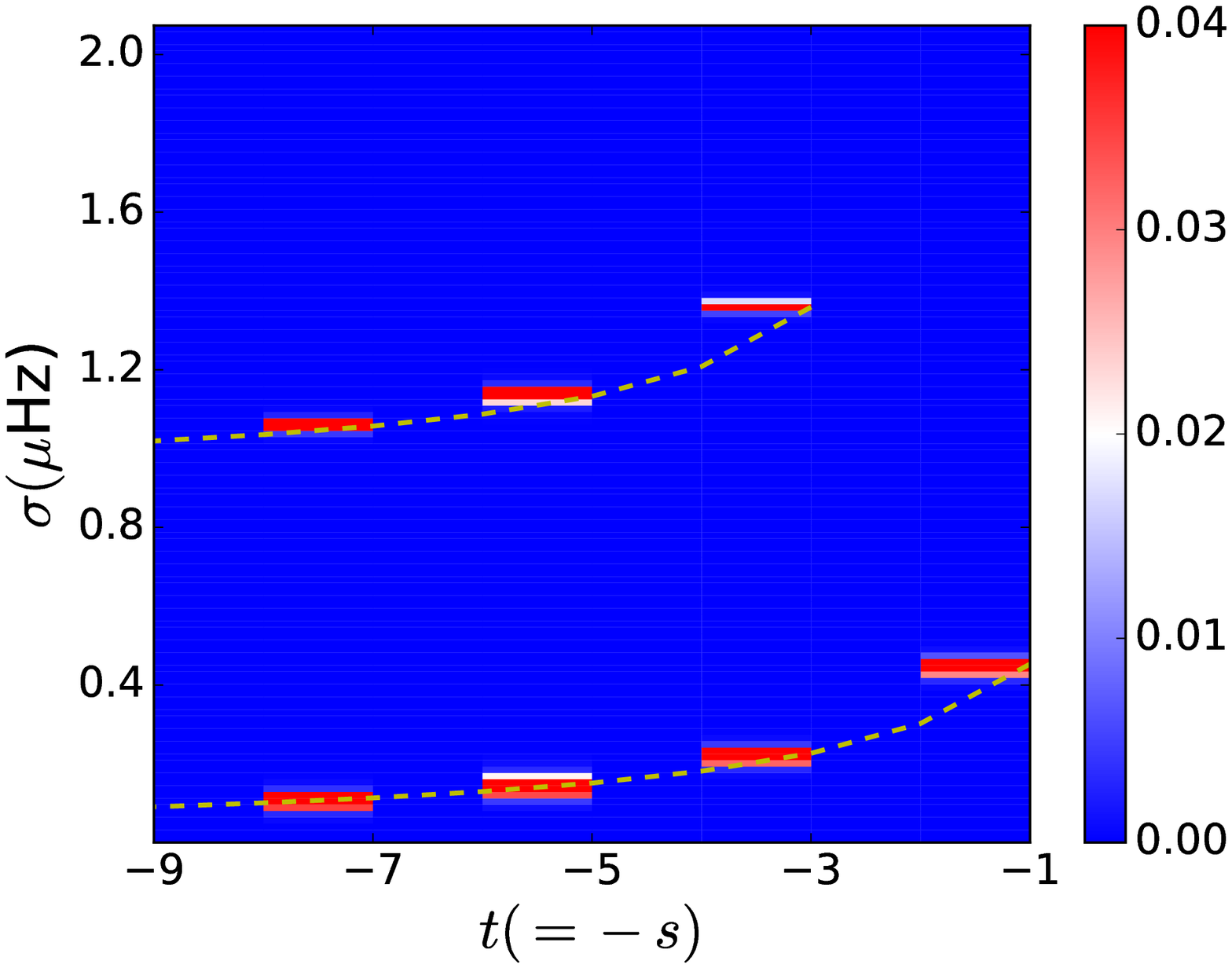}
\caption{\label{fig:leakage_inv_syn} Left panel displays leakage of modes when tracking is not applied. Leakage then occurs at the same frequency. The right panel displays leakage of modes when tracking is considered and bears a strong resemblance to observations, i.e. Figure~(\ref{fig:leakage_data}). In both cases, modes with odd harmonic degrees leak into neighbouring odd harmonic degrees.}
\par\end{centering}

\end{figure}

As explained in Appendix~\ref{sec:appendix}, the dispersion relation of Rossby waves either leaks into $\Omega$ or $2\Omega$ depending on the value of $L_{s}^{s^\prime}$. In order to appreciate which of $L_s^{s+1},$ or $L_s^{s+2}$ is more significant, we calculate $B$-coefficients using Equation~(\ref{eq:B_leak_ns}) without considering tracking and obtain the quantity $(\sum_{n}(-1)^{\ell}B^\sigma_{st}(n,\ell))^2$. From the left panel in Figure~(\ref{fig:leakage_inv_syn}), we see that power from odd $s$ leaks to odd $s$ and not to even $s$ for our measurements, which implies that $L_{s}^{s+2}$ is more significant than $L_s^{s+1}$. Because of this reason, we observe leakage at $\omega_t+2\Omega$ but not at $\omega_t+\Omega$. The right panel of Figure~(\ref{fig:leakage_inv_syn}) shows the effect of leakage if we track the data, similar to the spectrum observed in Figure~(\ref{fig:leakage_data}). 
There might be other systematics, e.g., P or B-angle corrections, that can affect the properties of leakage as discussed above. 

 \subsection{\label{sec:syn_inv}Synthetic inversions taking into account leakage}
 \citet{hanasoge18} have shown that leakage complicates the inversion using Equation~(\ref{eq:B_leak}). Therefore in our earlier work \citep{hanasoge19}, we performed inversions using Equation~(\ref{B_no_leak}) for simplicity's sake instead of the full Equation~(\ref{eq:B_leak}). This simplifying assumption might diminish the accuracy in retrieving the depth profiles of Rossby waves. In our synthetic test, we use cubic polynomials to characterize the depth profiles, with the condition that they are set to zero at depths $0.9R_\odot$ and below. We subsequently calculate the $B$-coefficient using Equation~(\ref{eq:B_leak}). In our first approach, we ignore leakage and assume $b^\sigma_{st}$ is same as $B^\sigma_{st}$ and then use Equation~(\ref{B_no_leak}) to invert for the profile $w^\sigma_{st}$. 
 
\begin{figure}
\begin{centering}
\includegraphics[scale=0.45]{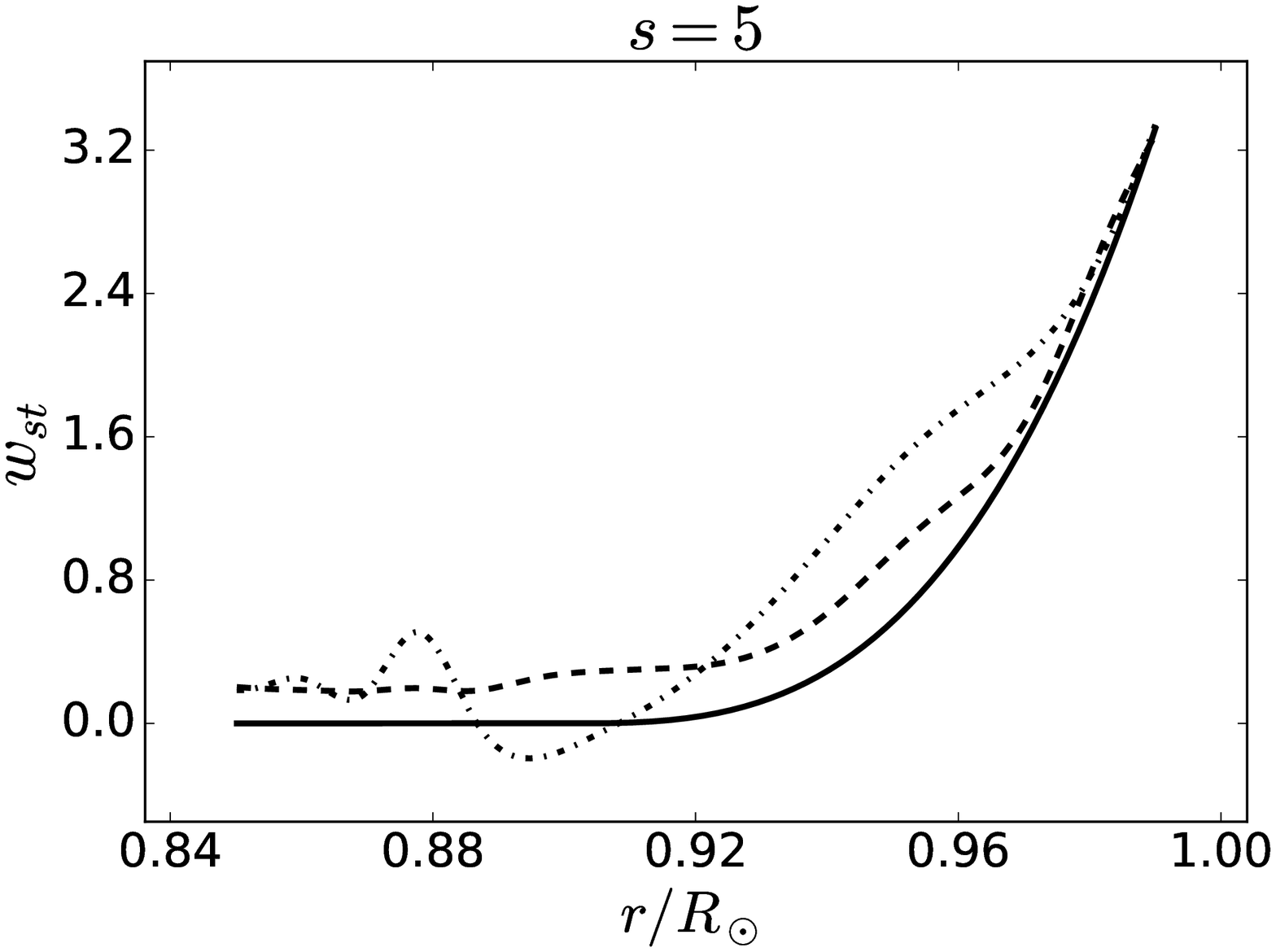}\includegraphics[scale=0.45]{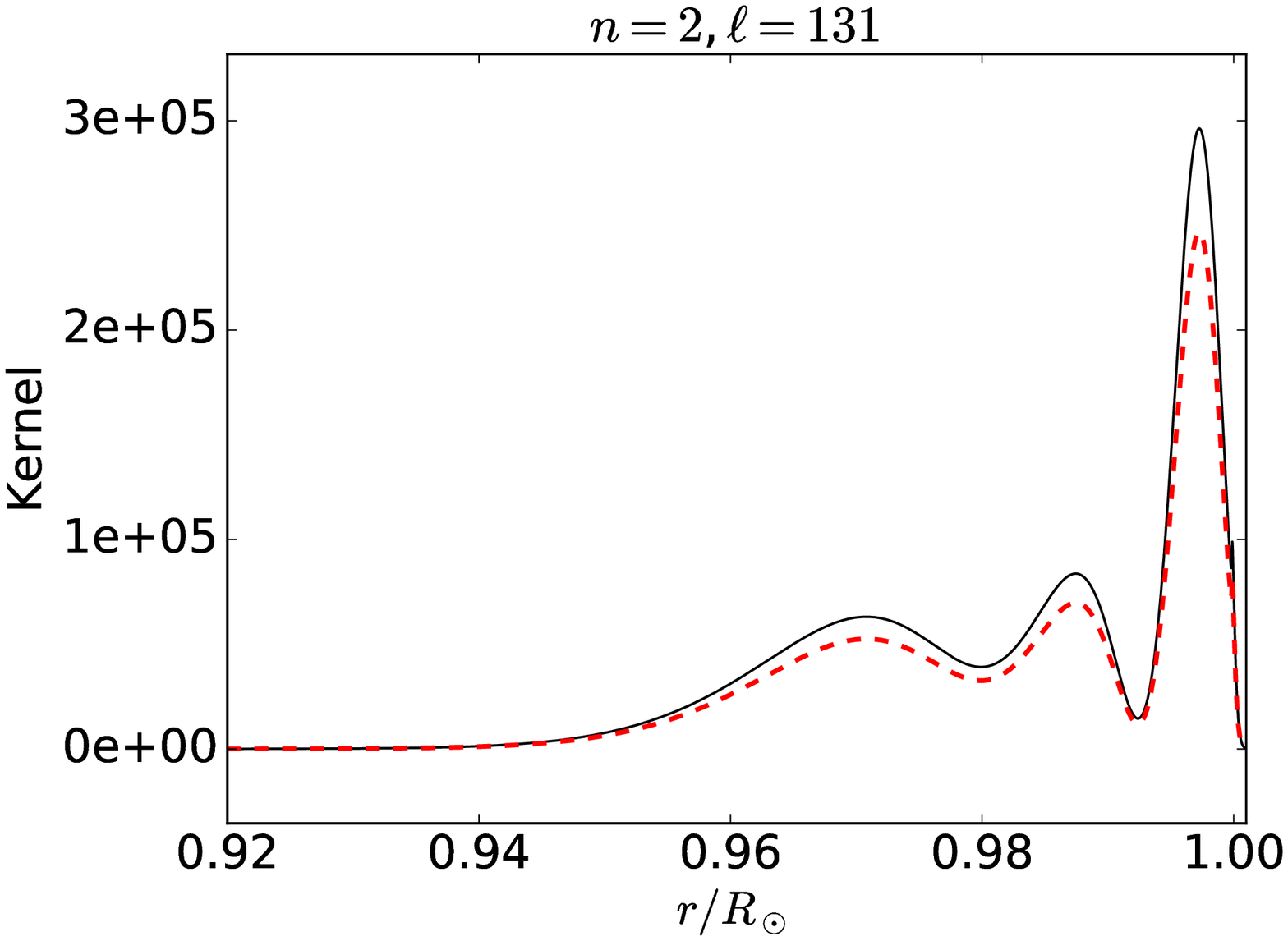}
\caption{\label{fig:kernel_comparison_synthetic_inv}{Left panel shows inversion results without noise. The input profile is plotted using a solid line and the inverted profile with and without leakage are marked by dashed and dot-dashed lines respectively. In the right panel we compare kernels, $f_{0,s}\mathcal{K}_{n\ell}(r)$ (red dashed line) with $\Theta_{s,-s}^{s,-s}$ (black solid line) for $s=7$. It can be seen that two kernels are of the same shape but slightly differing in magnitude from each other.
}}
\par\end{centering}
\end{figure}

\begin{figure}
    \centering
    \includegraphics[scale=0.7]{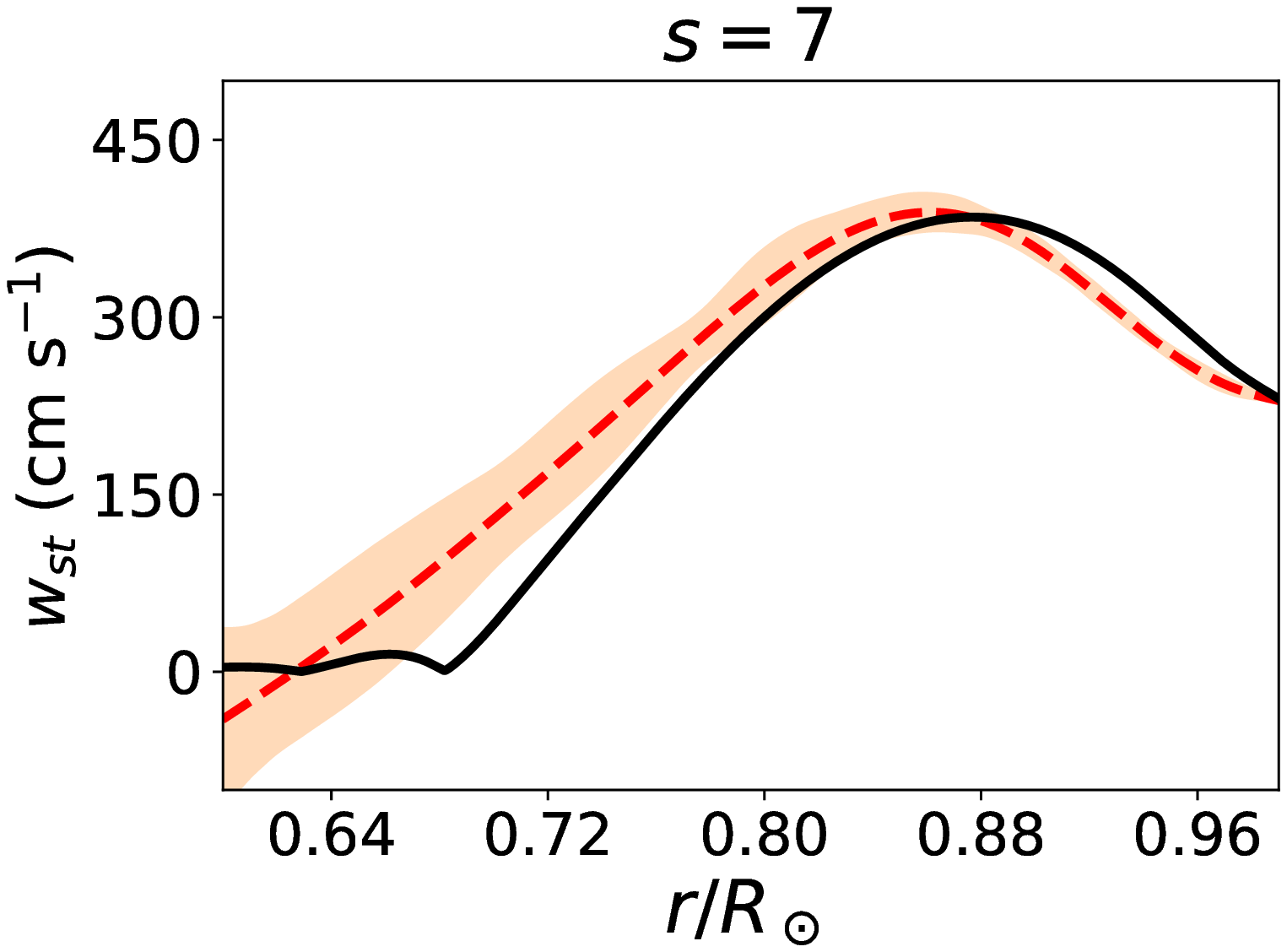}
    \caption{ Plot shows the inversion result with noise. In this case, we ignore leakage. The black solid line is the original profile we put in. The red dashed line is the inferred profile and corresponding error ( $\pm 1\sigma$ around the mean) in the inferred profile is shown by orange shaded area.}
    \label{fig:inv_noise}
\end{figure}

Next, we take leakage into account and proceed with the following approach. We write Equation~(\ref{eq:B_leak_ns}) using the following compressed form
 \begin{equation}
      B^\sigma_{s,-s}(n,\ell)=\int dr \sum_{s^\prime} \Theta_{s}^{s^\prime}
                (n,\ell,\sigma, r)w_{s^\prime -s^\prime}^{\sigma}(r), \label{B_theta}
 \end{equation}
 where 
 \begin{equation}
     \Theta_{s}^{s^\prime}(\sigma)=N^\sigma_{\ell s\,-s}\mathcal{K}(n,\ell)(r)\sum_{\ell^\prime,\ell^{\prime\prime},m,m^\prime,\omega}L_{\ell m}^{\ell^\prime,m^{\prime}}L_{\ell m-s}^{\ell^{\prime\prime}m^\prime-s^\prime}\gamma_{-s^\prime m}^{\ell s\ell}H^{\sigma *}_{\ell\ell m\,-s}
      \gamma^{\ell^{\prime\prime}s^\prime\ell^\prime}_{-s^\prime m^\prime}H^{\sigma}_{\ell^{\prime}\ell^{\prime\prime}m^{\prime}-s^{\prime}},\label{Theta_st}
 \end{equation}
 is the new sensitivity kernel which relates observed B-coefficients to properties of Rossby waves $w^{\sigma}_{s\,-s}$. The off-diagonal terms, $\Theta_{s}^{s^\prime}$ quantify contributions from neighbouring modes $(s^\prime,-s^\prime)$ to our desired mode $(s,-s)$ in the observed B-coefficient. The term $\Theta_{s}^{s^\prime}(\sigma)$ depends on the leakage matrix $L_{\ell m}^{\ell^\prime m^\prime}$. We have shown in section~(\ref{sec:leakage}) that leakage from neighbouring modes $s^\prime$ occurs at the same temporal frequency as that of the mode $\sigma_{s^\prime}$ if we do not employ tracking. If the data are tracked, this leakage moves to higher temporal frequencies. Since we are performing inversions at each frequency bin in the range $[0.0,0.5]\,\mu$Hz and since the contribution to our desired frequency bins from neighbouring modes is negligible in that range, we consider diagonal terms only, i.e. $\Theta_{s}^{s}$ in Equation~(\ref{Theta_st}) for the inversion. 
 \begin{equation}
    B^\sigma_{s,-s}(n,\ell)=\int dr  \Theta_{s}^{s}
                (n,\ell,\sigma, r)w_{s, -s}^{\sigma}(r).\label{diagonal}
 \end{equation}
 We apply Equation~(\ref{diagonal}) to invert for $B$-coefficients, estimated using Equation ~(\ref{eq:B_leak}). In the left panel of Figure~(\ref{fig:kernel_comparison_synthetic_inv}), we compare our inversion results with the original profile, demonstrating that choosing diagonal terms in Equation~(\ref{diagonal}) improves the inverted profile. This test also shows that choosing Equation~(\ref{B_no_leak}) is not a particularly bad assumption for this problem, and that our inferred amplitude might at worst be off by a factor of a few. It is the kernel that is responsible for  differences in the inferred amplitudes. To compare the two kernels, we plot $f_{0,s}\mathcal{K}_{n\ell}$ and $\Theta_{s}^{s}$ in the right panel of  Figure~(\ref{fig:kernel_comparison_synthetic_inv}) for a mode with radial order $n=2$ and harmonic degree, $\ell=131$. It can be seen in Figure~(\ref{fig:kernel_comparison_synthetic_inv}) that though the shape of the kernels remains same in two cases, the values are different, resulting in changes in the inferred amplitude. In order to test the inversion algorithm with noise added to the measurement, we choose an artificial profile that goes to zero at the base of the convection zone. The magnitude at the surface is set to $4$ ms$^{-1}$, which is close to the observed value. We calculate $B$-coefficients in a similar manner as described above using Equation~(\ref{eq:B_leak}) and add random Gaussian noise in proportion to the observed level, $\noise_{n\ell}^\sigma$. We then perform inversions assuming Equation~(\ref{B_no_leak}) and compare our inversion results with the original profile in Figure~(\ref{fig:inv_noise}). In order to get error in the inverted profile we repeat this process multiple times and estimate the standard deviation of these values, thereby obtaining the error in the inferential profile.


\subsection{\label{sec:freq_fit}Frequencies and line-widths of r-modes}
\begin{figure}
\begin{centering}
\includegraphics[scale=0.45]{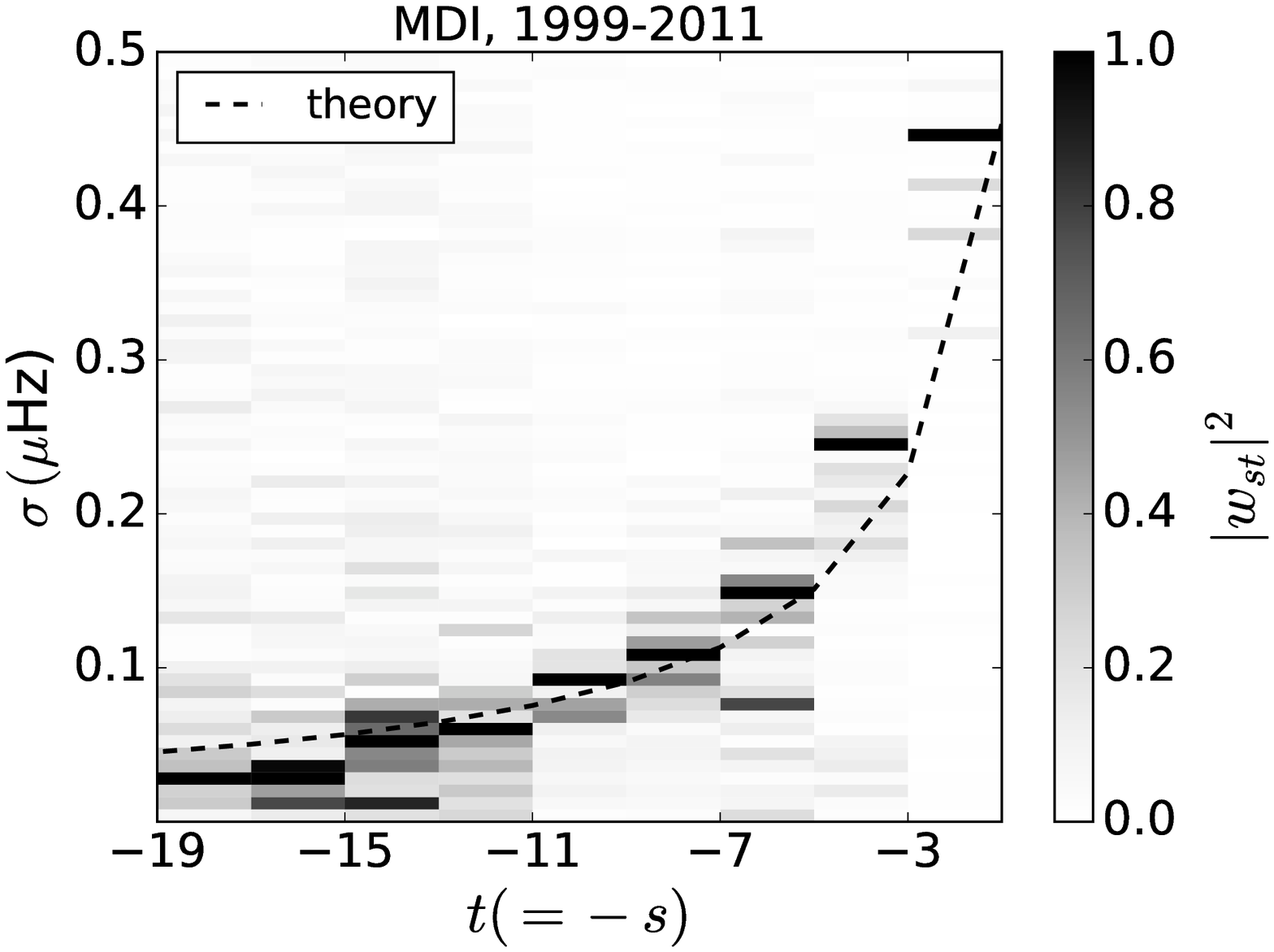}\includegraphics[scale=0.45]{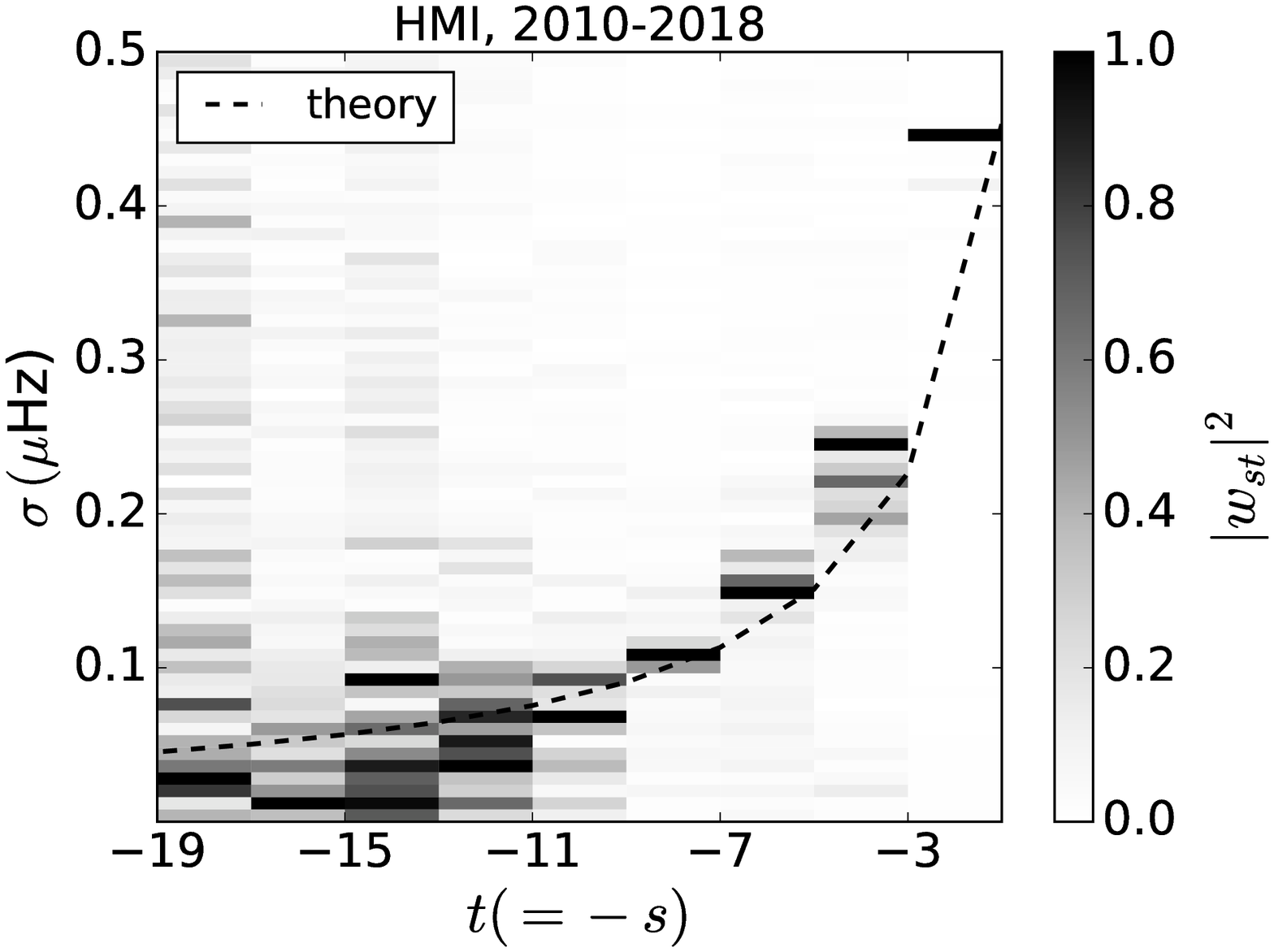}
\caption{\label{fig:HMI_MDI_power} The left panel shows the normalized average power spectrum of Rossby waves at depth $0.98 R_\odot$ by analyzing $12$ years of MDI data divided into three four-year chunks. The right panel shows the same as the left panel but with $8$ years of HMI data divided into two four-year chunks. The specific configuration of measurements we use allows us to only detect Rossby modes with odd harmonic degrees. The black dashed line in each panel represents the theoretical dispersion relation of sectoral Rossby modes in a co-rotating frame with rotation frequency $453$ nHz.}
\par\end{centering}
\end{figure}

\begin{figure}
    \centering
    \includegraphics[scale=0.5]{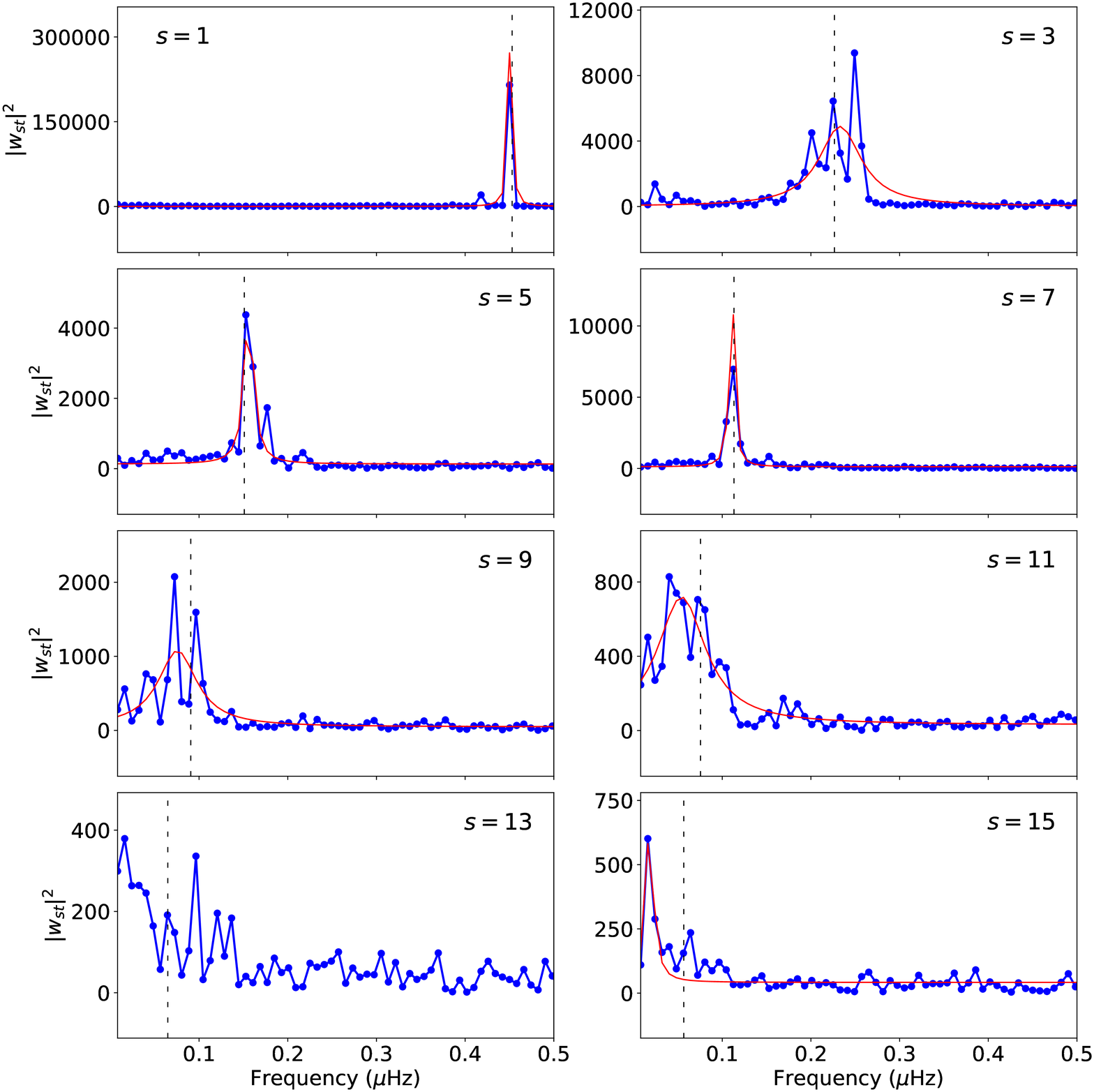}
    \caption{Averaged power spectrum from the analysis of SDO/HMI data (blue solid line with cycle). We fit a Lorentzian profile with a constant background to the power spectrum as described in section~(\ref{sec:freq_fit}). The fitted parameter values are tabulated in Table \ref{tab:freq_HMI}. The fitted spectrum is shown by red solid line.}
    \label{fig:HMI_freq}
\end{figure}

\begin{figure}
    \centering
    \includegraphics[scale=0.5]{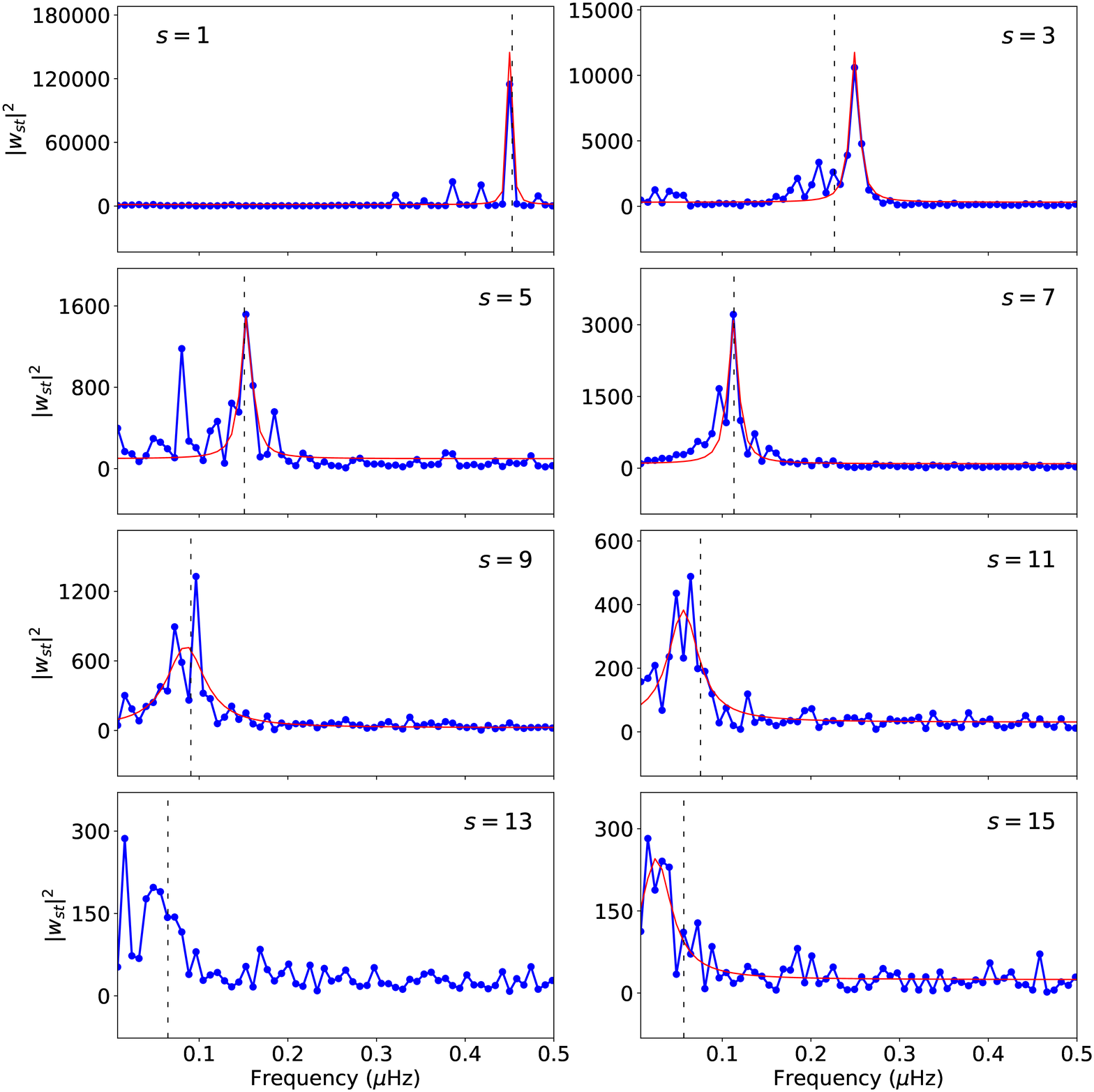}
    \caption{Same as in Figure~(\ref{fig:HMI_freq}). Averaged power spectrum (blue solid line with cycle) and corresponding fit (red solid line) from the MDI analysis. The fitted parameter values are listed in Table~(\ref{tab:freq_MDI}).}
    \label{fig:MDI_freq}
\end{figure}

After validating our method we estimate B-coefficients as described in section~(\ref{sec:data}) from the observed oscillation data. We divide $12$ years of MDI data into three four-years chunks and 8 years of HMI data into two four-years chunks. We analyze each chunk of data sets separately and do the inversion to obtain, $w^\sigma_{st}$ from the measured B-coefficients. We then take average of these results and plot the  average power spectra which is shown in Figure~(\ref{fig:HMI_MDI_power}). In order to quantify frequencies, line-widths and amplitude of these modes,  we fit a Lorentzian function plus a constant background 

\begin{equation}
    F(\omega)=\frac{A}{1+[(\omega-\omega_0)/(\tau/2)]^2}+B,\label{loretz}
\end{equation}
to $\vert w_{st}\vert ^2$ for each $s$. Here $A$ is the maximum amplitude of the Lorentzian, $\omega_0$ is the central frequency, $\Gamma$ is the full width at half maximum, $B$ is constant background power.  We use the {\it curve\_fit} module implemented in {\it scipy.optimize} to fit the power spectrum. We have tabulated values of these parameters for all modes starting from $s=1$ to $s=15$ obtained through analyses of HMI and MDI data in Tables~(\ref{tab:freq_HMI}) and~(\ref{tab:freq_MDI}) respectively. Fitted spectrum for HMI and MDI are shown in Figure~(\ref{fig:HMI_freq}) and~(\ref{fig:MDI_freq}) respectively. The $s=3$ mode parameters obtained from HMI data are similar to the values reported by \citet{gizon18}, whereas the MDI analysis is similar to the findings of \citet{liang_2018}.  The $s=13$ mode obtained from the analysis of HMI data does not clearly stand out and so we do not fit this mode. Amplitudes of all the modes are varying with harmonic degrees and becoming very small for harmonic degrees $s\ge 13$. Since our analysis period covers cycles $23$ and $24$, differences in parameter values between Tables~(\ref{tab:freq_HMI}) and~(\ref{tab:freq_MDI}) (obtained from analysis of HMI and MDI data respectively) may in principle carry information about the solar cycle dependence of Rossby modes. However, this interpretation is complicated because of our use of data from two different instruments that we have not cross calibrated. \citet{liang17} have performed this in the context of time-distance helioseismology. A similar task is required for normal mode coupling in order to combine MDI and HMI analyses.  \par  
{
\renewcommand{\arraystretch}{1.5}
\begin{table}[t]
\begin{center}
\setlength\tabcolsep{7pt}
\begin{tabular}{c c c c c c c c}
\hline
& &MC&\citet{gizon18}&\citet{liang_2018}\\
s&$\frac{2\Omega}{s+1}$&$\omega_{0}/(2\pi)$ &$\omega_{0}/(2\pi)$ &$\omega_{0}/(2\pi)$&$\Gamma/(2\pi)$& $\sqrt{A}$ & B\\
 &nHz&nHz&nHz&nHz&nHz&cm s$^{-1}$\\
\hline
1&453&451$\,\pm\,$0.1&—&—&5$\,\pm\,$0.5&538$\,\pm\,$157&31$\,\pm\,$647\\
3&226.5& 233$\,\pm\,$3&259&254$\,\pm\,$2&58$\,\pm\,$11&70$\,\pm\,$24&6$\,\pm\,$155\\
5&151&156$\,\pm\,$0.4&157$\,\pm\,$4&156$\,\pm\,$2& 12$\,\pm\,$1&68$\,\pm\,$15&131$\,\pm\,$22 \\
7&113&111 $\,\pm\,$0.1 &112$\,\pm\,$4&110$\,\pm\,$4&6$\,\pm\,$1&124$\,\pm\,$66&119$\,\pm\,$19\\
9&90.6&76$\,\pm\,$4&86$\,\pm\,$6&$82^{+4}_{-5}$&53$\,\pm\,$12&32$\,\pm\,$ 12&47$\,\pm\,$34\\
11&75.5&54$\,\pm\,$2&75$\,\pm\,$7&46$\,\pm\,$7&66$\,\pm\,$7&26$\,\pm\,$7& 31$\,\pm\,$12\\
15&56.6&18$\,\pm\,$1&$47^{+7}_{-6}$&$22^{+2}_{-3}$&10$\,\pm\,$3&25$\,\pm\,$12&42$\,\pm\,$6\\
\hline
\end{tabular} 
\caption{Analysis of HMI data. Mode frequency, $\omega_0$, amplitude $\sqrt{A}$, full width at half maximum $\Gamma$, background power $B$ that give best Lorentzian fits to the observed $B$-coefficient spectra are tabulated in the co-rotating frame. The second column in the table gives the theoretical frequencies of modes in a co-rotating frame with tracking frequency $453$ nHz. For comparison, we list the observed frequencies from other two studies, \citet{gizon18} and \citet{liang_2018}. The fitted spectrum is plotted in Figure~(\ref{fig:HMI_freq}). Here, MC stands for mode coupling.}
\label{tab:freq_HMI}
\end{center}
\end{table}
}
{
\renewcommand{\arraystretch}{1.5}
\begin{table}[t]
\begin{center}
\setlength\tabcolsep{7pt}
\begin{tabular}{c c c c c c c c}
\hline
& &MC&\citet{gizon18}&\citet{liang_2018}\\
s&$\frac{2\Omega}{s+1}$&$\omega_{0}/(2\pi$) &$\omega_{0}/(2\pi)$ &$\omega_{0}/(2\pi)$&$\Gamma/(2\pi)$& $\sqrt{A}$ & B\\
 &nHz&nHz&nHz&nHz&nHz&cm s$^{-1}$\\
\hline
1 &453& 451$\,\pm\,$ 0.1 &—&—& 5$\,\pm\,$1 &391$\,\pm\,$ 136 &1357$\pm$ 480 \\
3 &226.5&249$\,\pm\,$ 0.4 &259&254$\,\pm\,$2&  12$\,\pm\,$ 2 & 107 $\,\pm\,$ 37&306$\,\pm\,$ 72 \\
5 &151&153$\,\pm\,$ 1&157$\,\pm\,$4&156$\,\pm\,$2&15$\,\pm\,$ 3 &38$\,\pm\,$ 14 & 97$\,\pm\,$ 22\\
7& 113& 112$\,\pm\,$ 0.6 &112$\,\pm\,$4&110$\,\pm\,$4&14$\,\pm\,$ 2 &55$\,\pm\,$ 15 &93$\,\pm\,$ 24\\
9& 90.6&85$\,\pm\,$ 3&86$\,\pm\,$6&$82^{+4}_{-5}$&52$\,\pm\,$ 9 &26$\,\pm\,$ 9  &23$\,\pm\,$ 18 \\
11& 75.5&56$\,\pm\,$ 2 &75$\,\pm\,$7&46$\,\pm\,$7&  41$\,\pm\,$ 5 & 19$\,\pm\,$ 5  &30$\,\pm\,$ 6\\
13& 64.7&48$\,\pm\,$ 4&40$\,\pm\,$10&24$\,\pm\,$7&65$\,\pm\,$ 12&12$\,\pm\,$ 4&24$\,\pm\,$ 4\\
15& 56.6&25$\,\pm\,$ 2 &$47^{+7}_{-6}$&$22^{+2}_{-3}$&41$\,\pm\,$ 6&15$\,\pm\,$ 3&24 $\,\pm\,$ 4\\
\hline
\end{tabular} 
\caption{Same as in Table~(\ref{tab:freq_HMI}), except parameter values are obtained from analyzing MDI data. The fitted spectrum is shown in Figure~(\ref{fig:MDI_freq}).}
\label{tab:freq_MDI}
\end{center}
\end{table}
}    

 
 \subsection{Is the mode of degree $s=1$ due to systematics in the method?}
\begin{figure}
\begin{centering}

\includegraphics[scale=0.45]{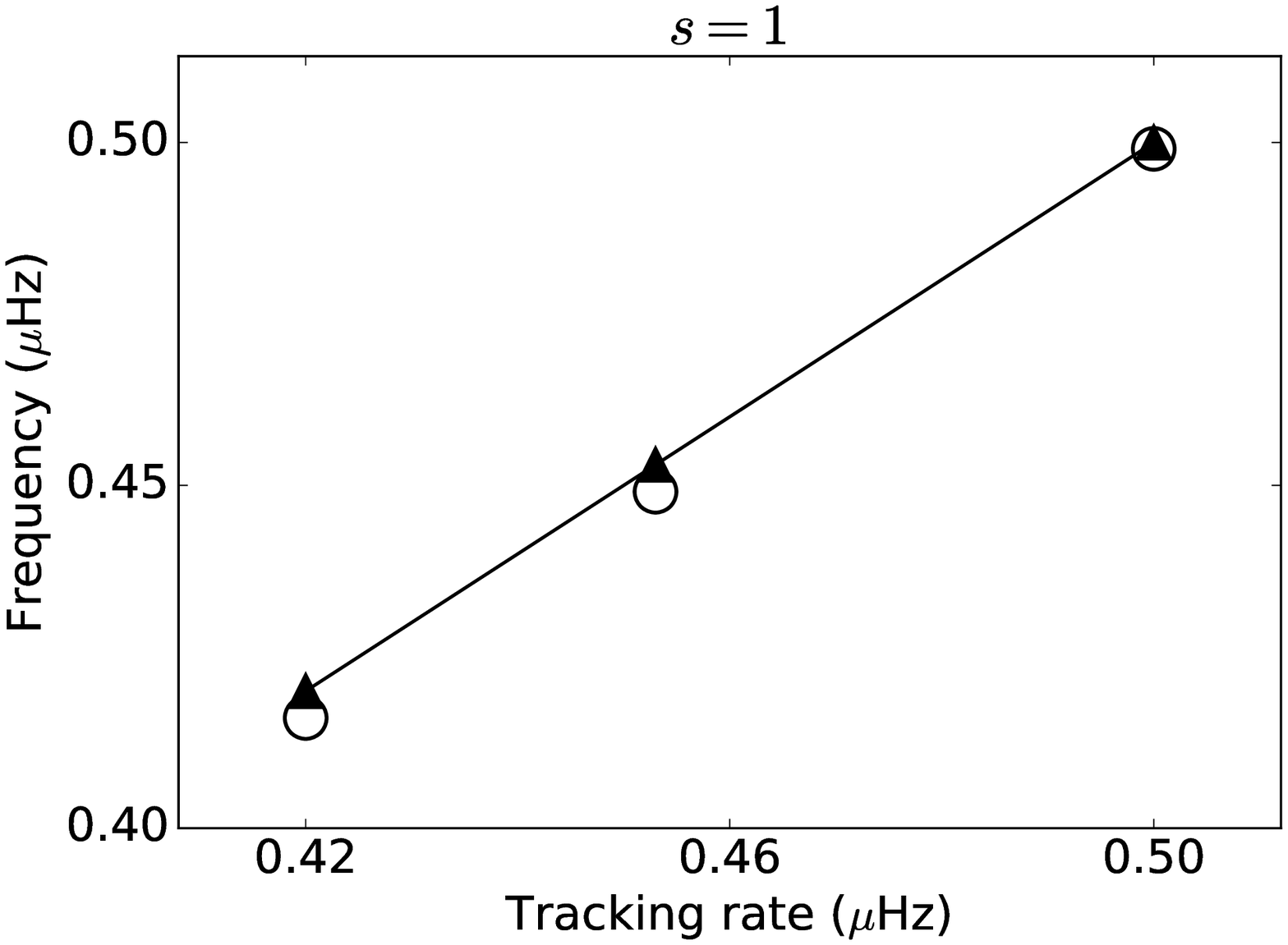}\includegraphics[scale=0.45]{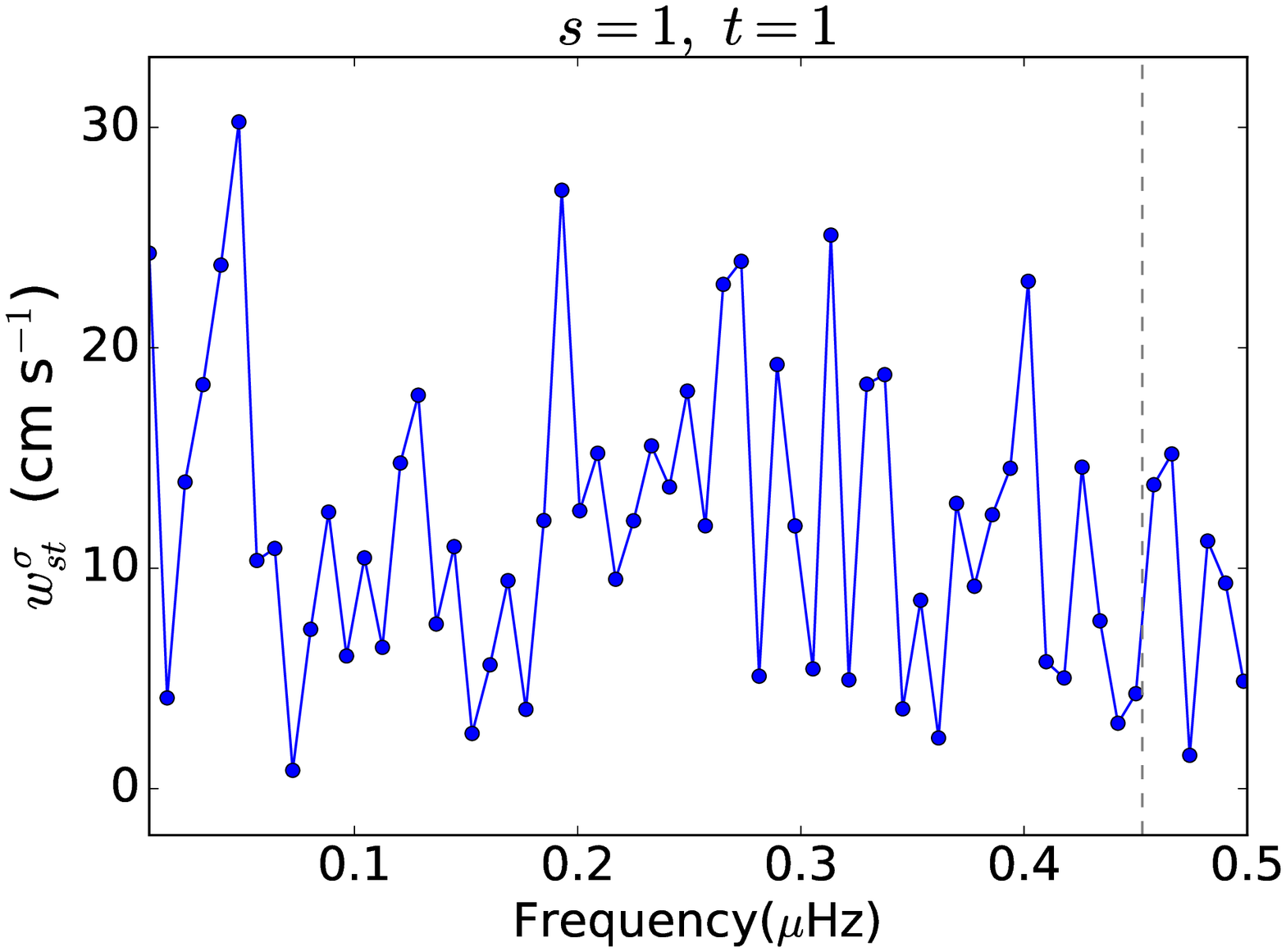}
\caption{\label{fig:s=1}Left panel:Frequencies of the mode $s=1$ (circles) for different tracking rates and corresponding theoretical values (triangle) are plotted. Right panel: we plot the power for harmonic degree, $s=1$ and azimuthal number, $t=1$. In the spectra, we do not find any extra power close to frequency $453$ nHz (dashed vertical line).}
\par\end{centering}
\end{figure}

 \citet{hanasoge19} detect the $s=1$ Rossby mode. Since the tracking rate is same as the frequency of the mode, $s=1$, systematics in our method might induce spurious power in the related spatial and temporal frequency bins. To investigate if it is due to tracking, we choose different values for it and calculate the B-coefficient for each case. If the tracking rate changes, so will the frequency of each mode in the power spectrum plot in Figure~( \ref{fig:HMI_MDI_power}).  We plot the frequency of the mode, $s=1$ versus tracking rate and compare it with the theoretical value in Figure~(\ref{fig:s=1}). We also see that the $s=1$ mode is leaking to mode $s=3$ in Figure~(\ref{fig:leakage_data}), which is unlikely to occur if it were to be due to systematics in the method. These arguments are in favour of the theory that the $s=1$ power is associated with Rossby waves and not due to systematics. We also plot the power for harmonic degree, $s=1$ and azimuthal number, $t=1$ in the right panel of Figure~(\ref{fig:s=1}). We do not find extra power close to the frequency $453$ nHz. Nevertheless, caution should be taken when interpreting this mode as we might not have accounted for all the systematics in our method.  One possibility might be centre-to-limb systematic which is viewed as an outflow from the disk centre to the limb. This is a poloidal flow. Nevertheless, if there is a component of this poloidal outflow that leaks to a toroidal component with harmonic degree, $s=1$ and azimuthal number, $t=-1$, spurious power at this temporal and spatial scale would show up in the spectrum. This fictitious flow is static from an inertial frame and therefore in a co-rotating frame its frequency will be equal to the tracking rate $\Omega$, which is same as the Rossby mode, $(s,t)=(1,-1)$. 

\section{Conclusion}
We have extended the work by \citet{hanasoge19} by analyzing $8$ years of HMI data by considering all observed oscillation modes of harmonic degree in the range $[10,180]$ to measure frequencies, line-widths and amplitude of Rossby modes. We also analyze $12$ years of  SOHO/MDI data from years, $1999$ to $2011$. In this work, we do not combine these two data sets, since a more sophisticated analysis to correct for P and B-angle variations is required \citep[as demonstrated by][in the context of time-distance helioseismology]{liang17}. Therefore, we show results from the two data sets separately. 

In order to understand how Rossby mode eigenfunctions vary with depth, we first validate our inversion technique by recovering a synthetic profile that we introduce. We first use Equation~(\ref{B_no_leak}) and perform the inversion. Next we consider the diagonal term in Equation~(\ref{B_theta}) for the inversion. 
In Figure~(\ref{fig:kernel_comparison_synthetic_inv}), we compare the inverted and original profiles. Choosing the diagonal term for the inversion provides a higher quality inference than when simply choosing Equation~(\ref{B_no_leak}). 

We performed tests to verify if the $s=1$ mode that we detect is due to systematics in our measurement since the frequency of this mode is same as tracking rate we choose, $\Omega$.  We choose different tracking rates and find that the observed frequency of the $s=1$ mode follows the theoretical prediction for that tracking rate. We also see that $s=1$ leaks to $s=3$ (see Figure~\ref{fig:leakage_data}), which is unlikely to occur if it were to be due to systematics in our method. However we remain skeptical and will not declare detection of the $s=1$ mode.\par
Our approach of modelling leakage and measurement technique differs from that of \citet{schad_2013}, who studied meridional circulation in the Sun using a variant of the measurement we apply here. Here, we consider coupling of identical-harmonic-degree p-modes, which limits us to the study of Rossby waves of odd harmonic degrees. In order to investigate the even harmonic degrees, we need to consider coupling between p-modes of different harmonic degrees, with $\delta\ell=1,3,\ldots$ (as $\delta\ell$ increases, frequency separation also increases, which in turn decreases the sensitivity of the coupling, as may be seen from Equations \ref{eq:B_coeff} and \ref{eqH}). Additionally, in order to image meridional circulation and differential rotation, we have to estimate $B$-coefficients as defined in Equation \ref{eq:B_coeff} at $\sigma=0$ (since we assume that these features only weakly evolve in time). We also mention in section \ref{sec:syn_inv} how the particular problem of Rossby waves in which we are interested becomes simplified because of its finite $\sigma$. Our future attempts will be to address more general problems in helioseismology using this technique.


\appendix
\section{\label{sec:appendix2} Partial to full sphere observation by ignoring leakage}
In case of full-sphere observations of the Sun, there is no leakage, i.e. $L_{\ell m}^{\ell^\prime m^\prime}=\delta_{\ell\ell^\prime}\delta_{mm^\prime}$. The general expression for the $B$-coefficient is 
\begin{equation}
    B^\sigma_{st}(n,\ell)=N^\sigma_{\ell st}\sum_{\ell^\prime,\ell^{\prime\prime},m,m^\prime,s^\prime
      ,t^\prime,\omega}L_{\ell m}^{\ell^\prime,m^{\prime}}L_{\ell m+t}^{\ell^{\prime\prime}m^\prime+t^\prime}\gamma_{tm}^{\ell s\ell}H^{\sigma *}_{\ell\ell mt}
      \gamma^{\ell^{\prime\prime}s^\prime\ell^\prime}_{t^\prime m^\prime}H^{\sigma}_{\ell^{\prime}\ell^{\prime\prime}m^{\prime}t^{\prime}}b^\sigma_{s^\prime t^\prime}(\ell^\prime,\ell^{\prime\prime}).\label{eq:B_leak_Appendix}   
\end{equation}
If we substitute $L_{\ell m}^{\ell^\prime m^\prime}=\delta_{\ell\ell^\prime}\delta_{mm^\prime}$ into Equation \ref{eq:B_leak_Appendix} we obtain
\begin{eqnarray}
    B^\sigma_{st}(n,\ell)=N^\sigma_{\ell st}\sum_{\ell^\prime,\ell^{\prime\prime},m,m^\prime,s^\prime
      ,t^\prime,\omega} \delta_{\ell\ell^\prime}\delta_{m m^\prime}\delta_{\ell \ell^{\prime\prime}}\delta_{m+t\,m^\prime+t^\prime}\gamma^{\ell s\ell}_{tm}H^{\sigma}_{\ell \ell m t}\gamma^{\ell^{\dprime}s^\prime\ell^\prime}_{t^\prime m^\prime}H^\sigma_{\ell^\prime\ell^\dprime m^\prime t^\prime}b^\sigma_{s^\prime t^\prime}(\ell^\prime,\ell^\dprime) \nonumber \\
     =N^\sigma_{\ell s t}\sum_{m s^\prime \omega}\gamma^{\ell s\ell}_{t m}\gamma^{\ell s^\prime \ell}_{t m}\vert H^{\sigma *}_{\ell \ell m t}\vert^2
     b^\sigma_{s^\prime t}(n,\ell).
     \label{eq:B_leak_a}
\end{eqnarray}
Note that we have put dependence on radial order $n$ back in the last step of the above equation. We currently observe only sectoral Rossby modes, i.e. $s=-t$ in the Sun. If we apply this condition to the above equation and using Equation \ref{eq:N_sigma}, we arrive at 
\begin{eqnarray}
    B^\sigma_{s\, -s}(n,\ell)=N^\sigma_{\ell s \,-s}\frac{1}{N^\sigma_{\ell s \,-s}}b^{\sigma}_{s\,-s}(n\ell),\nonumber\\
    =b^{\sigma}_{s\,-s}(n\ell).\label{eq:Btob}
\end{eqnarray}
Note that, to arrive at the final Equation \ref{eq:Btob}, we have assumed that only sectoral modes of Rossby waves exist.

\section{\label{sec:appendix}Spatial and temporal leakage}
We observe only a part of the solar disk that comes into the field of view of the telescope. If $v_{R}(\theta,\phi; T)$ is the velocity field due to Rossby waves, our measurement will be $v(\theta,\phi; T)$ and these two fields are related by the following expression 
\begin{equation}
   v(\theta,\phi;T)=W(\theta,\phi)v_{R}(\theta,\phi; T).\label{window}
\end{equation}
$W(\theta,\phi)$ is the window function unity over the visible part of the solar disk and otherwise zero. In our convention, $T$ denotes time and $t$ is used for azimuthal number. We assume that the velocity field for Rossby waves, $v_{R}(\theta,\phi; T)$, is scalar for simplicity. The velocity field of Rossby waves in the co-rotating frame may be written as 
\begin{equation}
    v_{R}(\theta^\prime,\phi^\prime;\omega^\prime)=\sum_{s,t}w_{st}(\omega^\prime)Y_{st}(\theta^\prime,\phi^\prime),\label{expnd}
\end{equation}
where $Y_{st}$ is the spherical harmonic with azimuthal number $t$ and harmonic degree $s$. $w_{st}(\omega)$ is the inferred toroidal velocity field at that spatio-temporal frequency bin and radius. The prime denotes the coordinates of a co-rotating frame  at angular frequency $\Omega$ rotating (with respect to the inertial frame). Coordinate and frequency transformations from one to other coordinate system are given by

\begin{align}
    \theta^\prime=\theta,
    \phi^\prime=\phi+\Omega T,
    \omega^\prime=\omega-t\Omega. \label{transform}
\end{align}
We omit the prime on $\theta$ in all equations below. Substituting Equation~(\ref{transform}) into~(\ref{expnd}) and using~(\ref{window}), we obtain 
\begin{eqnarray}
    v_R(\theta,\phi+\Omega T; T)=\frac{1}{2\pi}\int d\omega^\prime e^{i\omega^\prime T} \sum_{s}w_{st}(\omega^\prime)Y_{st}(\theta,\phi)e^{it\Omega T},\nonumber\\
    v_R(\theta,\phi;T)=\frac{1}{2\pi}\int d\omega e^{i\omega T}\sum_{s}w_{st}(\omega-t\Omega)Y_{st}(\theta,\phi).\label{r_velocity}
\end{eqnarray}
Using Equations~(\ref{r_velocity}) and~(\ref{window}), we have 
\begin{equation}
    v(\theta,\phi;T)= W(\theta,\phi)\frac{1}{2\pi}\int d\omega e^{i\omega T}\sum_{s}w_{st}(\omega-t\Omega)Y_{st}(\theta,\phi).\label{FFT_needed}
\end{equation}
 Performing a spherical-harmonic transform of Equation~(\ref{FFT_needed}), we obtain
 \begin{equation}
     v_{st}(T)=L_{st}^{s^\prime t^\prime}\frac{1}{2\pi} \int d\omega e^{i\omega T}w_{s^\prime t^\prime}
     (\omega-t\Omega), \label{temporal}
 \end{equation}
 where $L_{st}^{s^\prime t^\prime}$ denotes leakage from mode $(s, t)$ to $(s^\prime,t^\prime)$,  
 \begin{equation}
     L_{st}^{s^\prime t^\prime}=\int\sin\theta d\theta d\phi Y_{st}^{*}(\theta,\phi)Y_{s^\prime,t^\prime}(\theta,\phi) W(\theta,\phi). \label{L_st}
 \end{equation}
Access to the full-sphere observation would have meant $L_{st}^{s^\prime t^\prime} = \delta_{s s^\prime}\delta_{t t^\prime}$. After performing a temporal Fourier transform of Equation~(\ref{temporal}), we obtain
 \begin{eqnarray}
     v_{st}(\omega^\prime)=\int dT v_{st}(T)e^{-i\omega^\prime T},\nonumber\\
     = \frac{1}{2\pi}\int dT L_{st}^{s^\prime t^\prime}\int d\omega e^{i\omega T}w_{s^\prime t^\prime}
     (\omega-t\Omega),\nonumber\\
     =L_{st}^{s^\prime t^\prime} w_{s^\prime t^\prime}(\omega -t\Omega). \label{v_st_omega}
 \end{eqnarray}
Since we only detect sectoral modes of  Rossby waves, we drop $s$ or $t$ keeping in mind that $t=-s$. With this substitution, Equation~(\ref{v_st_omega}) becomes
 \begin{equation}
     v_s(\omega)=\sum_{s^\prime}L_{s}^{s^\prime}w_{s^\prime}(\omega-t^\prime\Omega). \label{drop_t}
 \end{equation}
 The frequencies of Rossby waves in the inertial frame follow the relation 
 \begin{equation}
     \omega=\sigma_t+t\Omega, \label{inertial}
 \end{equation}
 where $\sigma_t$ is the frequency of Rossby wave modes with azimuthal order $t$ which obeys Equation~(\ref{dispersion}). The power of these modes lies close to these frequencies (since the linewidths are small, see, e.g., Table~\ref{tab:freq_HMI}). For simplicity, we choose a delta function 
\begin{equation}
    w_s(\omega)=\delta [\omega-(\sigma_t+t\Omega)]. \label{power_ws}
\end{equation}
After substituting Equation~(\ref{power_ws}) into~(\ref{drop_t}), we obtain
\begin{equation}
    v_s(\omega)=\sum_{s^\prime}L_{s}^{s^\prime}\delta[\omega-(\sigma_t+t\Omega)].\label{vs_delta}
\end{equation}
At frequencies close to $\sigma_t+t\Omega$ we have 
\begin{equation}
    v_s(\sigma_t+t\Omega)=\sum_{s^\prime}L_{s}^{s^\prime}\delta[(\sigma_t+t\Omega)-(\sigma_{t^\prime}+t^\prime\Omega)],
\end{equation}
which is non zero only when $t=t^\prime$
\begin{equation}
    v_s(\sigma_t+t\Omega)=L_s^s\delta(0),
\end{equation}
which in turn is the classical Rossby-wave dispersion relation in an inertial frame. For frequencies $\sigma_{t+1}+(t+1)\Omega$ and $\sigma_{t+2}+(t+2)\Omega$, we recover
\begin{eqnarray}
    v_s(\sigma_{t+1}+(t+1)\Omega)=L_s^{s+1}\delta(0),\label{leak_1}\\
    v_s(\sigma_{t+2}+(t+2)\Omega)=L_s^{s+2}\delta(0)\label{leak_2}.
\end{eqnarray}
We would see power at frequency $\sigma_{t+1}+\Omega$ in a co-rotating frame with azimuthal order $t$ if $L_s^{s+1}$ were to be significant. Similarly, power at $\sigma_{t+2}+2\Omega$ is observed if $L_s^{s+2}$ were to be large.   

\bibliography{main}
 
\end{document}